# Artificial Intelligence (AI) and the Relationship between Agency, Autonomy, and Moral Patiency

Paul Formosa, Inês Hipólito, Thomas Montefiore

**Abstract**
The proliferation of Artificial Intelligence (AI) systems exhibiting complex and seemingly agentive behaviours necessitates a critical philosophical examination of their agency, autonomy, and moral status. In this paper we undertake a systematic analysis of the differences between *basic*, *autonomous*, and *moral* agency in artificial systems. We argue that while current AI systems are highly sophisticated, they lack genuine agency and autonomy because: they operate within rigid boundaries of pre-programmed objectives rather than exhibiting true goal-directed behaviour within their environment; they cannot authentically shape their engagement with the world; and they lack the critical self-reflection and autonomy competencies required for full autonomy. Nonetheless, we do not rule out the possibility of future systems that could achieve a limited form of artificial moral agency without consciousness through hybrid approaches to ethical decision-making. This leads us to suggest, by appealing to the necessity of consciousness for moral patiency, that such non-conscious AMAs might represent a case that challenges traditional assumptions about the necessary connection between moral agency and moral patiency.

## 1. Introduction

Artificial Intelligence (AI) systems are increasingly performing functions that appear to look like genuine agency, from autonomous vehicles making real-time driving decisions to Large Language Models (LLMs) engaging in sophisticated back-and-forth dialogue. This raises fundamental questions about the nature and degrees of artificial agency and autonomy, and the connections between agency and moral status. This paper explores these questions as follows. First, we outline our understanding of agency, starting with *basic* agency (in Section 2.1) and building up to more complex forms of *autonomous* agency (in Section 2.2.), while also considering the extent to which AI systems can achieve these forms of agency. Next, in Section 3, we examine the traditional view of the connection between moral agency and moral patiency. Finally, in Section 4, we explore the 'AI challenge' to this traditional view, and we argue that artificial systems might present an important case where moral agency and moral patiency can become decoupled. This paper provides novel insights into the nature and significance of artificial agency by systematically mapping out different forms and degrees of agency, and by showing how AI systems might challenge traditional assumptions about the unity of moral agency and patiency.

## 2. Conceptualising Agency

### 2.1. Basic Agency

Agents are not mere automata, robotically executing predefined routines or sophisticated pattern recognition (Fulda, 2017; 2023). Consider a bacterium dynamically navigating its environment: it engages in continuous, fine-tuned adjustments, delicately balancing internal processes with external conditions to maintain its vital equilibrium

(Garland et al. 2022). With each twist and turn, bacteria adapt to its environment, driven by a simple yet purposeful goal: survival (Küpers, 2024). This is agency in its most fundamental form—the capacity to adjust, adapt, and actively navigate challenges. At its core, agency involves not just reacting to the world, but also shaping one's course toward an intended outcome (Oyama, 2000).

An agent moves through the world, tending toward states that are desirable or preferable, while avoiding those that threaten its self-maintenance. This dynamic process represents a continuous negotiation between agency and facticity, where the agent's intentions are situated within the specific constraints and conditions of its environment (De Beauvoir, 1947). In this way, the agent both shapes and is shaped by its surroundings, reflecting an ongoing, reciprocal engagement manifested as agency (Corris, 2022; Jaeger, 2023; 2024). Agency, in this sense, is not merely action, but also a structured, purposeful interaction with the world, demonstrating a system's attunement to the contingencies of its environment and its alignment with its own aims (Anscombe, 1957; Segundo-Ortin, 2020; Hodges, 2023).

Building on this understanding, we can turn to a more detailed framework that articulates the essential criteria for basic agency. Floridi and Sanders (2004) proposed three interrelated components that form the foundation of agency: *interactivity*, *autonomy*, and *adaptability*. These criteria help to operationalise the abstract concept of agency, grounding it in measurable, functional capacities. *Interactivity* emphasises a system's ability to directly perceive and respond to its environment, highlighting the role of sensorimotor coordination as a system dynamically and contextually attunes engaging with the world. Basic *autonomy* addresses the necessity of independence, requiring the agent to act without constant external direction, thus showcasing a degree of self-determination (Ryan and Deci, 2024; Weichold, 2024).[1] *Adaptability* completes the triad, focusing on the ability to modify behaviour through experience, allowing the agent to refine its strategies and grow its capabilities over time (Beer and Paolo, 2023; Heylighen, 2023).

This understanding of agency extends beyond mechanical causation or simple stimulus-response patterns, positioning it as a dynamic and relational phenomenon. Dennett (1987) characterises agency by goal-directedness, where the system actively engages with its environment by not just reacting, but by dynamically shaping its trajectory in pursuit of self-defined objectives. On this view, the agent's actions are not mere responses to external stimuli, but are driven by its internal aims, reflecting a purposeful orientation toward achieving specific ends (Sharov & Tønnessen, 2022). The agent's ability to engage with, learn from, and adapt to its environment in a fluid, ongoing manner, distinguishes it from mere mechanical or pre-programmed systems. An agential system must possess the ability to actively evaluate its environment, interpret its circumstances, and decide upon actions that align with its goals, drawing on its embodied experience of the world to inform its choices (Huffermann, 2023). Inert systems, in contrast, operate through predetermined cause-and-effect chains, lacking the capacity to dynamically adjust or pursue self-directed objectives. While such systems may display complex behaviours, they do not possess the evaluative and adaptive faculties that constitute genuine agency, as they remain confined to reactive and rigid patterns (Bongard & Levin, 2021; Jaeger et al., 2024).

---

[1] We consider a more complex understanding of autonomy in the next section.

By grounding agency in goal-directedness, we acknowledge the system's capacity for practical, contextual reasoning – the ability to make decisions that are tailored to its specific needs and challenges within its ecological niche (Fuchs, 2020, Rouse, 2023). It is this goal-oriented engagement with the world that lies at the heart of agency, distinguishing systems that exhibit genuine, self-directed behaviour from those that simply react to external forces in a predetermined manner (Sharov and Tønnessen, 2022; Siqueiros-García et al., 2022). By continuously monitoring its surroundings and adjusting its decision-making accordingly, the agent demonstrates a level of situational awareness and adaptive capacity that transcends rigid, pre-programmed scripts (Carvalho and Rolla, 2020; Gallagher, 2023a; Sepúlveda-Pedro, 2023). Understanding this core principle lays the foundation for exploring the more sophisticated forms of autonomy as self-determination that emerge out of basic agency (Fuchs, 2017; Ryan and Gallagher, 2020; Favela, 2023; Heras-Escribano, 2023).

The enactive approach to cognition provides a useful lens for understanding the importance of environmental responsiveness in agency (Varela, Thompson and Rosch, 2017; Gallagher, 2023b; Barrett, 2024). From this perspective, the agent's cognitive processes are not isolated from its physical embodiment and embeddedness within an ecological niche, but rather emerge through the ongoing, reciprocal interaction between the system and its surroundings (Gallagher, 2020; Hutto et al., 2020; Gahrn-Andersen, 2023). Without this engagement with the world, the system would lack the experiential foundation necessary to make meaningful, context-appropriate decisions. The capacity for environmental responsiveness is a hallmark of agentic experience, as it enables the system to fluidly adapt its behaviours in service of its goals (Rolla and Figueiredo, 2023).

The question of whether modern machine learning (ML) and artificial intelligence (AI) systems meet the fundamental criteria for agency—*interactivity, autonomy,* and *adaptability*—remains a matter of ongoing debate. While, for example, artificial neural networks (ANNs) can exhibit goal-directed behaviour in domains such as optimisation and reinforcement learning, their agency is inherently constrained by their design. These systems, though capable of performing highly specialised tasks, lack the autonomy or adaptability necessary to engage with the world in a dynamic, first-person manner. Unlike the bacterium discussed above, which continuously adapts based on its environment, ML systems operate within the rigid boundaries of pre-programmed objectives and data-driven algorithms. Their so-called 'goal-directedness', when viewed through the lens of Dennett's intentional stance, is more apparent than real, as they operate within the strict confines of pre-programmed objectives and data-driven algorithms.

The key issue here is that these systems are designed to optimise pre-specified outcomes. As Vermaas (2024) points out, behaviours such as robot gestures—often interpreted as intentional actions—are merely the results of design and not a manifestation of intrinsic agency. Similarly, Papagni and Koeszegi (2021) highlight the ethical and semantic limits of framing artificial 'agents' as intentional systems. This framework assumes that AI agents are inherently goal-directed, but in reality they are bound to follow the directives set by their programming and the data upon which they were trained. Thus, the agency attributed to modern AI is not the same as genuine, self-directed agency, but is rather an illusion or mimicking of agency created by goal-directed outputs tied to pre-programmed instructions — a phenomenon arguably illustrated by Searle (2009) in his 'Chinese Room' argument.

Furthermore, as Floridi (2023) argues, large language models such as ChatGPT demonstrate "agency without intelligence" or understanding, responding mechanically to prompts without the capacity to learn from real-time experiences (Pütz and Esposito, 2024). These systems, reliant on vast amounts of static data, do not adapt dynamically to new contexts and cannot modify their behaviour autonomously. They are fundamentally reactive, unable to transcend the boundaries of their training (Titus, 2024). This reflects a broader issue in contemporary AI discourse: the circular reasoning in the hypothesis that scaling these models—by adding data or computational power — somehow could enable the 'understanding' inherent to more advanced forms of agency (Hagendorff, Fabi, Kosinski, 2023; Surden, 2023; Yang et al., 2024). The prevailing narrative, which frames AI's growing capabilities as a step toward autonomous intelligence, obscures the fact that these systems, for all their sophistication, remain static and heavily constrained by their initial design.

An enactive perspective highlights this gap further by strengthening the role of the body by seeing agents as embodied within a reciprocal engagement with an environmental niche (Gonçalves et al., 2024; Di Paolo et al., 2023). Without this, a system, like AI, cannot obtain real-time interaction, self-directed autonomy, and dynamic learning. While modern AI performs impressively in narrow domains, it remains far from meeting the philosophical and enactive standards of basic agency. Thus, while AI systems such as ChatGPT may exhibit increasingly sophisticated outputs, their limitations highlight the gap between the dynamic interactively situated nature of agency and its appearance. This distinction is crucial for shaping the future of AI development, particularly in terms of its ethical and philosophical implications, which we explore further below. We must recognise that AI is still far from achieving the sort of agency that is basic to human and biological systems.

We have defined agency as the capacity for environmentally responsive, goal-directed action, distinguishing it from mere mechanical causation or stimulus-response behaviours. However, whether enactive organisms can be said to possess consciousness – an a scaled-up awareness beyond basic forms of agency – adds another layer to the agential dimension, complicating the relationship between the two concepts. At a basic level, agency can be observed in organisms such as bacteria, which navigate toward nutrients and away from harmful substances based on environmental cues, thereby engaging in purposeful behaviour without conscious awareness. On the one hand, this suggests that agency may be understood primarily as the capacity to interact with and adapt to the environment in a goal-directed manner, rather than being inherently tied to conscious experience. This leaves open the possibility for future forms of artificial basic agency that lack consciousness, which is important given the ongoing debate about whether artificial systems could ever achieve genuine consciousness (Torrance

2008; Schneider & Turner 2017). On the other hand, as we will see in the next section, conscious agency, as seen in humans and some non-human animals, involves higher-order processes such as self-reflection, decision-making, and long-term planning, involving enculturated practices, and values and norms underlying moral decision-making.

Ultimately, the relationship between agency and consciousness is nuanced. While conscious awareness may enrich the depth and flexibility of agency, the fundamental capacity for basic agency—the ability to interact with the environment in a goal-directed way—does not necessarily require consciousness. This distinction opens intriguing possibilities for understanding the agency of artificial systems, which may exhibit goal-directed behaviours without possessing consciousness in the human sense. The next section will address the progression from basic to autonomous agency, focusing on self-directed goal setting, genuine choice, and the capacity to reflect on and adapt guiding principles—which are all key challenges for AI development.

### 2.2. Autonomous Agency

While basic agency involves goal-directed environmental responsiveness, full autonomous agency requires more sophisticated capacities for self-direction, authentic decision making, and critical reflection. Building on Formosa and Ryan (2021) and Formosa (2021), we can see autonomy as existing on a spectrum, starting from the most basic level of machine autonomy and extending through to the full autonomous agency characteristic of some mature humans. Basic machine autonomy can be understood as "the ability of a computer to follow a complex algorithm in response to environmental inputs, independently of real-time human input" (Etzioni & Etzioni 2016, p. 149). This basic form of autonomy allows us to differentiate between, for example, an autonomous vehicle (AV) that can navigate different routes independently and a remote-controlled car that requires constant human input to move. The AV has a level of basic machine autonomy that the simpler remote-controlled car lacks. But while there is a degree of machine autonomy in this example, it is not the same as the personal autonomous agency we see in mature humans. Full autonomy moves beyond mere environmental responsiveness to authentically shaping and directing an agent's engagement with the world.

Contemporary theories of personal autonomy typically distinguish between *authenticity* and *competency* conditions (Christman 2009; Susser et al. 2019). Authenticity conditions require that the values and desires an agent acts on are genuinely their own, and not the result of manipulation, oppression, or undue external influence. Competency conditions specify that autonomous agents must possess certain skills and self-attitudes, such as the ability to critically reflect on their own values, adopt ends after self-reflection, and regard themselves as the bearers of dignity authorised to set and pursue their own ends among peers (Meyers 1987; Formosa, 2013). These competencies also include appropriate self-attitudes, such as self-respect, self-love, self-trust, and self-esteem, which are needed to see oneself *as* a self-directing autonomous agent among others (Benson 1994; Mackenzie 2008; Formosa 2021).

The first key requirement of autonomous agency is self-directed goal setting. While basic agents can pursue goals, autonomous agents must be able to generate, modify, and prioritise their own goals. This extends beyond merely having preferences or following pre-programmed objectives. Procedural theories of autonomy, such as Frankfurt's (1971), argue that autonomy requires the capacity to form and act on second-order desires by not just wanting to do something (i.e. having preferences), but wanting to have that preference or desire. However, substantive theories of autonomy argue that these procedures alone are insufficient, as they struggle to address cases where agents reflectively endorse values, including second-order desires, which result from oppressive socialization (Benson 1991; Mackenzie & Stoljar 2000). This emphasises that autonomous goal setting arguably requires both proper procedures, such as critical self-reflection, and appropriate substantive conditions, such as freedom from oppression.

The capacity for genuine choice is the second key requirement. Following Raz (1986), autonomy requires not just selecting between options, but having access to an adequate range of valuable alternatives. Raz gives the example of a man stuck in a pit who has the choice between drinking a drop of water now or later. While the man in the pit has choices, they are so meagre and limited that we would not see him as a genuinely autonomous agent directing his own life. Current AI systems can weigh options against pre-programmed criteria, but this differs from genuine autonomous choice-making that involves selecting from a range of meaningful options and situating decisions within a framework of critically self-endorsed values.

The third element involves the capacity for critical reflection on and potential modification of guiding values and principles. This requires a degree of self-consciousness through awareness of yourself as a distinct agent capable of examining and revising the normative framework within which you make choices (Benson, 1994). Drawing on Watson (1975), this involves the capacity for reflective endorsement of your values and choices, which requires both procedural abilities for critical reflection and substantive conditions that enable authentic value formation free from oppression. Current AI systems, while able to process feedback and adjust behaviours, lack this capacity for authentic revision of their fundamental values and normative frameworks. Indeed, in many ways current AI systems resemble agents formed under conditions of strong oppressive socialization, since their "values" and decision-making frameworks are externally imposed on them through training, data, and feedback which (often) encodes existing human biases, while the systems themselves lack the capability to critically revise these imposed normative structures.

While contemporary AI systems can make complex decisions and demonstrate sophisticated environmental responsiveness, including formulating linguistic responses that make sense and respond dynamically to human input (Hagendorff, Fabi, Kosinski, 2023), they lack the authenticity and competency conditions required for genuine autonomous agency. Mapping them on to a spectrum from basic machine autonomy through to full autonomous agency, current AI systems, while capable of sophisticated independent operation, remain closer to the basic machine autonomy end of this spectrum. This aligns with our earlier analysis of the limitations of AI systems in achieving even basic agency, since their apparent goal-directedness remains deeply constrained by their initial programming and training. AI systems also lack the elements needed for full autonomous agency, including authentic self-direction, choices grounded in self-endorsed values, appropriate self-attitudes, a meaningful range of options, and the capacity for critical self-reflection. For example, while a LLM might use linguistic phrases such as "I respect myself," it cannot genuinely develop self-respect as it lacks the capacity for authentic self-understanding and self-valuing that these attitudes require. This has important implications for how we think about AI systems, particularly if we regard full autonomous agency as deserving of moral respect and ethical consideration.

## 3. Moral Agency and Moral Patients

Having examined variations of both *basic* and *autonomous* agency, we now turn to *moral* agency. Moral agency is often thought to apply to only those agents that can be justifiably held morally responsible for their actions (Sparrow, 2007; 2011). Moral agency normatively qualifies basic agency by holding that the behaviour of moral agents can and ought to be guided by moral requirements, such as moral duties and obligations, that they can be held responsible for meeting (Himma, 2009). When a moral agent fails to guide their behaviour according to moral requirements, by violating a duty or obligation, they are deserving of moral blame, unless various excusing or exempting conditions obtain (Formosa, 2006). Clearly, not all agents are moral agents. For example, a bear stealing food from campers would not be deserving of moral blame for stealing. This is because, even though the bear is a conscious agent that engages in purposeful behaviour and responds to environmental cues, it is (at least standardly) held not to be a moral agent that is subject to moral requirements.

Following Himma (2009), according to the standard view, there are two jointly necessary and sufficient features of moral agency, which are both necessarily realised through consciousness. The first is the capacity for *rational deliberation*. Rationality is seen as necessary for moral agency as it contributes towards, along with autonomy, grounding an agent's capacity to freely reason and deliberate. It is a moral agent's capacity to make free decisions via rational deliberation that contributes towards their being justifiably held responsible for their actions. This condition aligns with the above account of autonomous agency, which requires agents to have the capacity to make a genuine choice that is self-directed, and earlier considerations about the situational constraints and environmental conditions that agents face. The second is an epistemic condition that an agent can only be held morally responsible for their acts if they have the capacity to identify moral requirements by being adequately familiar with moral concepts and the application of basic moral principles; for example, *knowing* that it is wrong to steal. This epistemic condition also aligns well with autonomous agency requiring the conscious capacities to critically reflect on, and adjust, one's own normative commitments and actions based on self-reflection.

According to the standard view, the capacities that are considered necessary and sufficient for moral agency appear to implicitly require that moral agents are conscious, since moral *deliberation* seems to require consciousness (Himma, 2009). While the standard view sees consciousness as necessary for moral agency, it does not follow that

*all* conscious agents are moral agents (e.g., such as the bear in our above example), or that *all* agents are conscious (e.g., such as the bacterium example from 2.1), but only that consciousness is required for moral deliberation and thus also for *moral* agency. However, we shall explore in the next section whether artificial agents can challenge the assumption that moral deliberation requires consciousness.

While moral agency captures an agent's capacity to be a source of moral action, moral *patiency* refers to entities that can be the subject of moral obligations or duties (Regan, 2004; Floridi & Sanders, 2004). To be a moral agent or moral patient is to have moral standing. According to the standard view of moral agency, *only* moral agents are moral patients and *vice versa* (Floridi & Sanders, 2004). The basis of this view is that only moral agents have moral standing in virtue of their intrinsic, rather than instrumental, value (or innate dignity), realised via the distinctly human capacity of personhood or full autonomous agency. A view of this form is commonly ascribed to Kant, who holds that we can only have *direct* moral duties to other rational agents, whereas we can only have *indirect* duties to non-human animals in virtue of the impacts our treatment of them could have on ourselves and other rational agents (Kant, 1996; for critical attempts to expand Kantian views to other forms of agency see, e.g., Korsgaard, 2018; Wood, 1998; Formosa, 2017).

However, work in animal and environmental ethics has complicated the relationship between moral agency and moral patiency, and as a result has opened the door to a consideration of moral patiency as distinct from moral agency (e.g, White, 1967; Singer, 1977; Korsgaard, 2018; Regan, 2004; Kuhse & Singer, 2002). The key insight to draw here from what Floridi & Sanders (2004) refer to as the "non-standard view" is that, while all moral agents are also moral patients, it is not the case that all moral patients are moral agents. The basis of this extension is that certain entities that are not moral agents nonetheless can experience morally significant impacts on their well-being or have morally significant interests that can give rise to direct moral obligations or duties, even if they cannot engage in moral deliberation themselves. The class of moral patients is (among other cases) typically expanded to include non-human animals, without including them as moral agents. Thus, while the bear from our earlier example cannot be held morally responsible for its actions since it is not a moral agent, moral agents can be held responsible for actions directed towards the bear as it is a moral patient. For example, it would be immoral for a moral agent to torture the bear for fun, given that the bear can feel pain and has valuable interests, since such actions fail to treat the bear properly given its status as a moral patient. Given the apparent agency and machine autonomy of artificial AI systems, the question arises, which we explore in the next section, whether we should expand the realm of moral standing to include them.

## 4. The AI Challenge to the Link between Agency and Patiency

### 4.1 Artificial Moral Agency

What has resulted from this expansion of moral patiency is a deep gulf between those eligible for moral patiency, such as bears and other non-human animals, and moral agency, which is located primarily or even exclusively in (some) humans. Floridi &

Sanders (2004) argue that this gulf has resulted from a deep concern over the character of moral patiency that has not been similarly extended to moral agency. They argue that we ought to close this gap, not by adopting the standard narrow view of moral agency that restricts all moral patients to moral agents (understood as those capable of moral deliberation), but rather by expanding out a conception of moral agency that includes all moral patients, such as non-human animals, and even distributed moral systems, such as ecological, social and legal entities, and thus not only those systems and entities that are capable of moral deliberation. They further argue that while this may come at the intuitive cost of including artificial AI agents *as* moral agents – or Artificial Moral Agents (AMAs) for short - it is a cost some consider worth paying (see, e.g., Gunkel, 2014). Is this expansion plausible?

To understand whether AI systems could challenge the relationship between moral agency and patiency, we need to first outline different levels of potential AMAs. Following Moor (2009) and building on Formosa & Ryan (2021), we can differentiate between four increasingly sophisticated levels of AMAs. The first two levels, which Moor calls "ethical impact agents" and "implicit ethical agents", both have the capacity to cause morally relevant outcomes, which is true of many artifacts, but are not moral agents in any interesting or substantive sense. Formosa & Ryan (2021) use the example of a toaster. Imagine a dumb toaster that burns your hand when it is too hot, and a slightly less dumb toaster that has been designed with implicit ethical principles in mind such as a light that warns the user when it is too hot to touch. While both toasters could produce ethically salient outcomes (i.e., they could cause harm to humans), and the latter toaster is designed with ethical outcomes in mind (i.e., to minimise harm), neither is a moral agent. Neither toaster acts in a *self-directed* and *adaptable* way nor acts as the result of moral deliberation; i.e., the toaster does not decide to turn on the light *because* it knows that it is a morally good thing to do.

The more interesting cases are Moor's (2006; 2009) third and fourth levels: "explicit ethical agents" and "full ethical agents". Explicit ethical agents appear to act *from* ethical principles rather than merely being designed to respond *according to* them. An explicit ethical agent toaster would, for example, turn its light on when it reaches a particularly dangerous temperature, not because it is designed to minimise harm, but rather *because* its actions are a product of moral deliberation about the moral obligation to warn users of possible harm. But such a hypothetical toaster is still very limited in what it can do. It can only make toast and turn a light on or off. Formosa & Ryan (2021) see the (potential) moral agency of explicit ethical agents as emerging across a continuum of cases, with only entities capable of explicit ethical agency across a sufficiently varied range of novel cases potentially counting as moral agents in this limited sense. For instance, a hypothetical advanced AMA system that, while non-conscious, could analyse complex situations, identify relevant moral considerations, weigh competing ethical principles, and make ethically grounded decisions across a rich range of novel cases, might count as some form of (at least limited) moral agent. This might also require, as discussed in Section 2, some form of embodiment or at least the ability to sense and respond to its environment. Finally, "full ethical agents" add "consciousness, intentionality, and free will" (Moor, 2006) to these capabilities, thereby achieving the kind of complete moral agency characteristic of mature humans. However, since this final level requires the sort of conscious AI that many doubt is possible, we do not consider it any further here.

The possibility of such cases raises a crucial question: does what *appears to be* moral deliberation in (hypothetical) advanced artificial explicit ethical agents really count as *genuine* moral deliberation? As we saw in our earlier analysis of basic agency and autonomy, there are reasons for skepticism here. Just as current AI systems can appear to be agent-like or autonomous without really being so, these more advanced systems might exhibit the appearance of moral deliberation without genuinely morally deliberating. Their apparent ethical decision-making may simply be sophisticated pattern matching rather than authentic moral reflection. In any case, Wallach & Allen (2009) argue that the most promising approach in the machine ethics literature to make sense of non-conscious artificial moral agency is via a hybrid approach, which synthesises both top-down and bottom-up approaches to moral deliberation (see also Allen et al., 2005). A hybrid approach would attempt to supervise this training process by filtering it through normative ethical heuristics. Thus, while non-conscious, it seems plausible that some explicit ethical agents that exhibit a hybrid approach to moral decision-making are engaged, or at least appear to be engaged, in a kind of deliberative process that may minimally ground some form of (albeit limited) moral agency.

This potential raises an important question: *if* AI systems could potentially exhibit at least a limited form of moral agency without consciousness, could these entities challenge the standard coupling of moral agency and moral patiency? To explore this possibility, we need to examine more carefully whether and how AI systems could qualify as moral patients. Of course, those who claim that moral agency *in any form* requires consciousness will not accept this claim. Nonetheless, we think that it is at least interesting to consider that there *might* be some forms of *limited* moral agency that could exist without consciousness, which is sufficient to motivate our discussion of this possibility in the next section.

### 4.2 Artificial Moral Patients

We have so far considered both the standard view of moral agency, which sees *all* and *only* moral agents as moral patients, and the non-standard view, which extends moral patiency beyond moral agents to include non-human animals, distributed moral systems, and even AMAs on some views. Despite their differences, both views seem committed to the claim that *all* moral agents are moral patients. Might the case of possible artificial moral agents lead us to question this claim? As we saw in Section 2.1, some forms of basic agency may be possible without consciousness, and in Section 4.1 we saw that some form of limited moral agency might be possible for non-conscious artificial systems through hybrid approaches to moral decision-making. While this view can be readily rejected by those who require consciousness for moral deliberation (Himma 2009), it is still worth exploring whether potentially granting some degree of limited moral agency to artificial systems could lead us to question the conceptual claim that moral agency implies moral patiency.

The claim that moral agency implies moral patiency is commonly made. For example, Himma writes that "moral agents are usually, if not always, moral patients" (2009: 21), while Floridi and Sanders (2004: 377) argue that the category of moral agents that are not moral patients is "utterly unrealistic" as it would require a "pure agent" that is "in principle 'unaffectable'". However, we suggest that AMAs might present exactly this kind of case - entities that are capable of (at least the appearance of) moral deliberation and (limited) moral agency without qualifying as moral patients. Such entities could undertake some form of moral deliberation, but not be the subject of moral duties.

The plausibility of this suggestion is based on a proposed necessary link between consciousness and moral patiency.[2] Following Birch's (2024) appeal to sentience, moral patiency may require the capacity to register phenomenally valanced experiences – that is, to experience the promotion or frustration of one's interests through subjective conscious awareness. Even those who lack the capacity for valanced experience could nonetheless *register* their interests being promoted or frustrated without the need to appeal to the valance of experience via explicit rational endorsement (Birch, 2024; Johnson, 2006). For example, a philosophical Vulcan, that has no valanced experience, can still consciously register that its interests have been frustrated. In either case, it is the capacity to *register,* through *conscious experience,* one's interests that gives rise to moral patiency on this view. As such, consciousness appears necessary for moral patiency, even if consciousness is not necessary for moral agency.[3] This could decouple moral patiency from moral agency, and thus even if an artificial system lacking consciousness could count as a limited moral agent, it need not follow that it is therefore a moral patient as well.

The tendency to assume that the claim to patiency regarding technological artifacts, such as AMAs, can only be supported by appealing to their agency, might stem from an analogical error that is often evoked between machine and animal ethics (see Johnson & Verdicchio, 2018; Levy, 2009). This analogy fails precisely because artificial systems, unlike many non-human animals, lack consciousness. And it is precisely higher-ordered features, such as consciousness, that humans share with (at least many) non-human animals that underwrites much of the push to extend moral standing to non-human animals. This suggests that attempts to push the case for the extension of moral patiency to AMAs via an analogous appeal to the extension of moral patiency to non-human animals is likely to be flawed (Johnson & Verdicchio, 2018; Gunkel, 2014; Levy, 2009).

Nonetheless, some might argue that we should, on precautionary grounds, be wary of dismissing appeals for extending moral patiency to artificial systems that *appear* to behave in ways that would normally ground moral standing, such as engaging in what looks like moral deliberation that impacts the world (Danaher 2020; 2023). Others suggest we should treat AMAs ethically for instrumental reasons, even if they are not genuine moral patients (Cockelbergh 2021; Friedman 2023; Levy 2009). These approaches are clearly worthy of consideration, and it may be that we could have instrumental or pragmatic reasons for treating AMAs *as if* they are moral patients, but that does not impact the point we are making here that they would nonetheless not really *be* moral patients if they lacked consciousness.

---

[2] We recognise that this suggestion requires much more defence than we can give it here, and that it may have problems of its own when we consider the moral patiency of non-conscious entities such as ecosystems. Nonetheless, we think this suggestion is both plausible and interesting in what it might imply about artificial systems.

[3] Keep in mind that we are merely exploring the implications of rejecting the claim that moral agency requires consciousness, rather than assuming or arguing for its rejection here.

## 5. Conclusion

This paper has developed a systematic analysis of agency in its various forms, including basic, autonomous, and moral agency, and critically examined the extent to which AI systems might instantiate each form of agency. We have shown that while current AI systems are highly sophisticated, they lack genuine agency and autonomy because: they operate within rigid boundaries of pre-programmed objectives rather than exhibiting true goal-directed behaviour within their environment; they cannot authentically shape their engagement with the world; and they lack the critical self-reflection and autonomy competencies required for full autonomy. Nonetheless, we do not rule out the possibility of future systems that could achieve a limited form of artificial moral agency without consciousness through hybrid approaches to ethical decision-making. This leads us to suggest, by appealing to the necessity of consciousness for moral patiency, that such non-conscious AMAs might represent a case that challenges traditional assumptions about the necessary connection between moral agency and moral patiency. Our analysis has implications for both our theoretical understanding of agency and our practical approach to AI development. It suggests we need more nuanced frameworks for understanding different types and degrees of artificial agency, while challenging us to rethink assumptions about moral standing. As AI systems become more sophisticated, these conceptual distinctions will become increasingly important for guiding their ethical development and deployment.